\newcommand{\be}{\begin{equation}}
\newcommand{\rn}{Reissner-Nordstr\"om}
\newcommand{\ee}{\end{equation}}
\newcommand{\bea}{\begin{eqnarray}}
\newcommand{\eea}{\end{eqnarray}}
\newcommand{\gbar}{\overline{g}}
\newcommand{\har}{\overline{h}}

\documentstyle[epsf,prd,aps]{revtex}
\begin{document}
\draft
\twocolumn[
\title{Black hole singularities: a numerical approach}
\author{Patrick R. Brady\protect\thanks{Address after October 1, 1995:
Theoretical Astrophysics,  California Institute of Technology, Pasadena,
California 91125} and John D. Smith}
\address{Department of Physics, The University, Newcastle Upon Tyne NE1 7RU}
\date{4 April 1995}
\preprint{NCL95-TP**}

\maketitle

\typeout{ABSTRACT}
\widetext
\begin{abstract}
\parbox{\textwidth}{\hfil\parbox{6in}{
The singularity structure of charged spherical collapse  is studied by
considering the evolution of the gravity-scalar field system.    A detailed
examination of the geometry at late times strongly suggests the validity of
the mass-inflation scenario~\cite{PI:90}.   Although the area of the
two-spheres remains finite at the Cauchy horizon,  its generators are
eventually focused to zero radius.    Thus the null, mass-inflation singularity
{\em generally}\/ precedes a crushing $r=0$ singularity deep inside the black
hole core.   This central singularity is spacelike.
}\hfil}
\end{abstract}
\hskip\baselineskip
]
\narrowtext

The radiative tail of gravitational collapse decays as an inverse power of time
leaving behind a black hole characterized only by its mass, charge and angular
momentum~\cite{MTW}.   The simplicity of the external field at late times
contrasts with the highly dynamical interior which lies beneath the event
horizon~\cite{Hiscock,PI:90,Ori:91,Ori:92}.

It is generally accepted that the ``tunnel'' through a charged or rotating
black
hole is destroyed by perturbations which propagate into the hole after its
formation~\cite{Penrose1,Chandra}.  The manner in which this occurs
has
been studied in great detail using simplified
models~\cite{Hiscock,PI:90,Ori:91,Ori:92},  and the results suggest that the
Cauchy horizon (CH) inside a
charged (or rotating) black hole is transformed into a singularity at which the
``{Coulomb}'' component of the Weyl curvature diverges.  The null generators of
the CH contract slowly under transverse irradiation, so it is anticipated that
this null singularity eventually gives way to a central spacelike singularity
deep inside the black hole core.  It should be noted that this spacelike
singularity is {\em not} expected to precede the CH as some authors have
argued~\cite{Yurtsever,Gnedin}.  Indeed, geodesic observers falling into
the black hole at late times will generally encounter only the null
singularity.

Some disquiet exists regarding this picture of the black hole interior.
Yurtsever~\cite{Yurtsever} has argued, on general grounds, that complete
destruction of the CH should be expected once generic perturbations are
considered.  He bases his discussion on experience with colliding plane wave
spacetimes --- he has shown that plane wave Cauchy horizons are replaced by
spacelike singularities in the presence of generic plane-symmetric
perturbations~\cite{Yurtsever:pw}.  Gnedin and Gnedin~\cite{Gnedin}, on the
other hand, have performed a numerical integration of the spherical
Einstein-Maxwell-scalar field equations.  They established the existence of a
spacelike singularity inside a charged black hole coupled with scalar matter.
Their analysis stressed the behaviour of the central ($r=0$) singularity,
however they did not consider the possibility that it intersects the Cauchy
horizon.

This letter reports on an independent numerical investigation.  We
find evidence that the analytic models demonstrate the essential features of
black hole internal structure.

The general spherical line element can be written as
	\begin{equation}
	ds^2 = - g\overline{g} dv^2 - 2 g dv dr + r^2 (d\theta^2 + \sin^2 \theta
			d\phi^2 ) \; , \label{eq:1}
	\ee
where $g = g(r,v)$, $\gbar = \gbar (r,v)$  and $v$ is advanced time.   We
solve the field equations $G_{\alpha\beta} = 8\pi (E_{\alpha\beta} +
T_{\alpha\beta})$  where
	\be
	E{^{\alpha}}_{\beta} = (e^2/8\pi r^4) \; \mbox{diag} (-1, -1, 1, 1)
	 \label{eq:2}
	\ee
is the standard electromagnetic contribution to the stress-energy, and
	\be
	4\pi T_{\alpha\beta} =   \phi_{,\alpha} \phi_{,\beta}
			- \frac{1}{2} g_{\alpha\beta} \left(\phi_{,\gamma}
			\phi^{,\gamma}\right) \label{eq:3}
	\ee
is the stress energy for a massless,  minimally coupled scalar field $\phi$.
It
is convenient to write $\har = \phi$ and then introduce the scalar $h(r,v)$ by
	\be
	(r\har)_{,r} = h\; . \label{eq:4}
	\ee
The field equations can now be written as
	\bea
	(\ln g )_{,r} &=& r^{-1} (h- \har)^2\; , \label{eq:5}\\
	(r\gbar)_{,r} &=& g(1 - e^2/r^2)\; ,\label{eq:6}
	\eea
while the wave equation $\Box \phi =0$ becomes
	\be
	h_{,v} - \frac{\gbar}{2} h_{,r} = \frac{1}{2r} (h-\har)(g[1 - e^2/r^2]
		- \gbar)\; .\label{eq:7}
	\ee

The pioneering numerical integration of these equations,  with $e=0$, was
performed by Goldwirth and Piran~\cite{Goldwirth}.  We have employed a similar
algorithm.  Using the method of characteristics Eq.~(\ref{eq:7}) is recast as a
set of $2n$ ordinary differential equations,  where $n$ is the number of radial
grid points.   The radial integrals in Eq.~(\ref{eq:4})-Eq.~(\ref{eq:6})
are discretized according to the trapezium rule,
while we use a Runge-Kutta scheme for the ordinary differential equations.
The
resulting code is locally second order accurate,  and has been tested on \rn\
spacetime and the exact self-similar solutions in~\cite{Japs}.   The
details will appear elsewhere.

To aid the comparison to previous work~\cite{PI:90,Ori:91} we introduce the
mass function
	\bea
	m(x^{\alpha}) &=&  \frac{r}{2} \left( 1 + e^2/r^2 - \gbar/g \right)
	\; .\label{eq:11}
	\eea
Moreover the sole non-vanishing Newman-Penrose Weyl scalar
	\be
	- \Psi_2 =  \frac{1}{2} C{^{\theta\phi}}_{\theta\phi} =
	[m(x^{\alpha} ) - e^2/r]r^{-3} \; ,\label{eq:11a}
	\ee
is simply expressed in terms of this mass function and provides a direct
measure of curvature near the CH.

Characteristic initial data are supplied on a pair of intersecting null
hypersurfaces as shown in Fig.~1.  The spacetime is \rn\ when $v < v_0$.
After this advanced time we consider the evolution of a charged black hole with
infalling
scalar matter.   Since our primary interest is the black hole interior it is
useful to choose the outgoing null hypersurface, $\Gamma$,  coincident with the
event horizon.  This is achieved by first setting $g|_{\Gamma} = -1$, then
specifying $r|_{\Gamma}$ such that
	\be
	\lim_{v\rightarrow \infty} r|_{\Gamma} = \mbox{constant}\; .
	\label{eq:8}
	\ee
\begin{figure}
\leavevmode
\vskip1cm\par
\hskip-1.5cm
\hbox{\epsfxsize=8cm \epsfysize=8cm {\epsffile{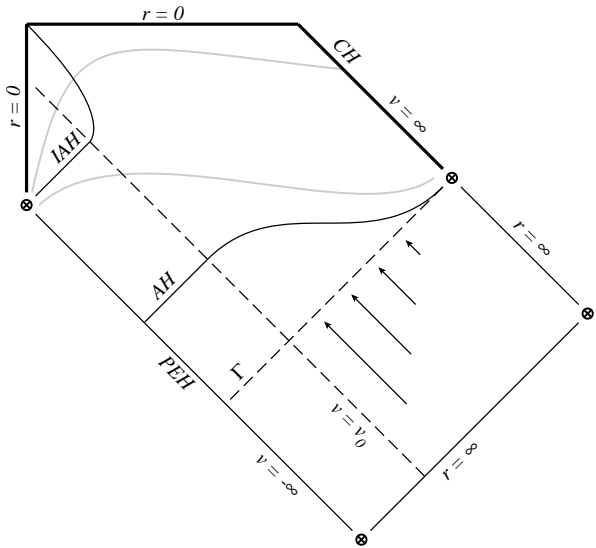}}}
\vskip-1cm\par
\caption{\label{figure1} A spacetime diagram showing the setting of the
numerical integration.  The spacetime is \rn\ for $v<v_0$.  Scalar field falls
into the black hole across the event horizon, $\Gamma$.  $PEH$ is the past
event horizon, located at $v=-\infty$.  $AH$ is the outer apparent horizon,
and $IAH$ is the inner apparent horizon of the charged black hole.  A couple of
lines of constant $r$ are shown in light gray.  The Cauchy horizon ($CH$) is a
singular hypersurface which contracts to meet the central singularity at $r=0$.
 All singularities are indicated by thick lines in the diagram.}
\end{figure}
\noindent
The dynamical equations (\ref{eq:4})-(\ref{eq:7}) are supplemented by one
further equation along each outgoing null hypersurface:
	\be
	\frac{d\gbar}{dv} = \gbar \left(\frac{d\ln g}{dv} -
	\frac{1}{2r^3} \left[g(r^2 - e^2) - \gbar\right]\right) + 2r\left(
	\frac{d\phi}{dv}\right)^2\; .
	\label{eq:9}
	\ee
This is Raychaudhuri's equation for the outgoing null rays,  and is conserved
by the evolution.  Using the characteristic equation $dr/dv = - \gbar/2$ to
obtain $\gbar|_{\Gamma}$, it is straightforward to integrate Eq.~(\ref{eq:9})
along $\Gamma$, with the appropriate initial
conditions at $v = v_0$, for $\phi|_{\Gamma}$.

Beneath the event horizon of the black hole outgoing null rays contract rapidly
(see Fig.~2).  Consequently an initially uniform
radial grid becomes highly non-uniform at late times, leading to significant
errors in $\gbar$.  We have employed a non-uniform grid along
the initial ingoing null ray to alleviate this problem, making the radial grid
points more dense in the
vicinity of the event horizon.  A significant increase in (accurate)
integration time is obtained by this method, and the internal mass function
[defined in Eq. (\ref{eq:11})] reaches values up to about $10^{50}$ times the
external black hole mass before errors become significant.

We have integrated the equations for several choices of initial data along
$\Gamma$,  and have obtained the same qualitative behaviour.
Table~\ref{table:1}
summarizes the results obtained for the two initial data sets characterized by
	\be
	r|_{\Gamma} =
	\left\{ \begin{array}{l}
	r_+ - \beta v^{-p} \; \\
	r_+ - \beta \exp{(-pv^2)}
	\end{array} \right.
	\label{eq:10}
	\ee
where $\beta>0$ and $p>0$ are real parameters.  It is convenient to fix
dimensions
such that the asymptotic Bondi mass is unity,  thus the radius of the event
horizon approaches $r_+ = 1 + \sqrt{1 - e^2}$ as $v\rightarrow \infty$.   The
inverse power-law data
along $\Gamma$ is taken as representative of the physical situation where
the flux of radiation across the event horizon exhibits such a
decay~\cite{Price}.  We have included the second data set to
emphasize that the non-linear instability of the inner horizon is
present even for perturbations having compact support on the event
horizon.  This result should be expected based on linear analysis of the
problem~\cite{Chandra}.

\noindent\parbox{3.5in}{
\begin{table}
\caption{\label{table:1}
Late time behaviour of the metric along characteristics labelled by $u$.
Multiplicative factors which depend only on $u$ have been omitted.  The
constants $\gamma$, $\mu$, and $\sigma$ are discussed in the text.
}\vskip0.1in
\begin{tabular}{cccc}
Initial data & $g(u,v)$ & $\gbar(u,v)$ & $m(u,v)$
\\ \tableline \\
$r_+ - \beta v^{-p}$ & $e^{-\gamma v}$ & $ v^{-(p+1+\sigma)}$  &
$v^{-(p+1 + \sigma)}e^{\gamma v}$\\
$r_+ - \beta e^{(-pv^2)}$ & $e^{-\gamma v}$ & $e^{-2 \mu v} $  &
$e^{(\gamma - 2 \mu) v}$\\
\end{tabular}
\end{table}}

In \rn\ spacetime all outgoing null geodesics which originate inside the black
hole intersect the CH within a finite affine parameter.  However, the
presence of the scalar field modifies the causal structure of the
spacetime so that some outgoing null rays terminate at $r=0$ in a finite
coordinate time $v$ (i.e. before intersecting the CH).   This important result
is demonstrated in Fig.~2, which shows the radius as a function of external
advanced time along a selection of outgoing null rays.  In this coordinate
system the lightcones tip over at $r=0$ indicating that the singularity at this
hypersurface is spacelike.

\begin{figure}
\leavevmode
\vskip-3.5cm
\hskip-0.5cm
\hbox{\epsfxsize=8cm \epsfysize=10cm {\epsffile{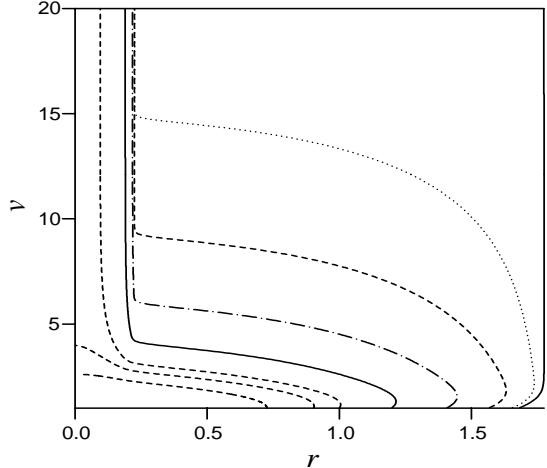}}}
\caption{\label{figure2}  The null rays in the $rv$-plane.  The inner-most rays
terminate at $r=0$ in a finite advanced time,  however there exist many
geodesics which approach a finite radius at large $v$.  Subsequent figures show
$\gbar$ and $g$ along the five outermost rays indicated by different line
types. The initial data is exponential with $\beta =  0.26$, $p = 1$ and $e^2 =
 0.4$.}
\end{figure}

Generally we have found the outgoing null rays to belong to two distinct
classes,  those which reach $r=0$ and those which do not.  The latter geodesics
are of greatest interest to us here and below.  Each member of this class
approaches a fixed radius at late times,  indicating the existence of a CH in
these solutions.  Notice that the final radius is different for each geodesic;
the CH is not a stationary null hypersurface,  as it is in \rn\ spacetime,
rather it contracts slowly under transverse irradiation.  Moreover, the CH is
the locus of a null, precursory singularity  which precedes the crushing
singularity at the origin (see Fig.~1).  Gnedin and Gnedin examined only the
central singularity in~\cite{Gnedin},  overlooking the existence of the null
portion.

One might worry that the cumulative effect of infalling scalar field could
eventually focus all outgoing null rays to zero radius.  To allay such doubts
we have examined the expansion rate $\gbar = -2\, dr/dv$ along outgoing null
rays.  Typically $\gbar$ reaches a maximum along each geodesic,  then it
exhibits a period of exponential decay followed by a less pronounced, but
definite, approach to zero which is characteristic of the initial data (see
Fig.~3).   Table~\ref{table:1} gives the form of $\gbar$ for large $v$,
demonstrating that the assumptions used in asymptotic calculations are
valid~\cite{Ori:92,Chambers}.

Our initial conditions along $\Gamma$ reflect the relaxation of the external
gravitational field to \rn\ spacetime at late times.   It seems reasonable that
this property should continue to hold inside the event horizon as far as some
imprecisely defined radius,  where non-linear gravitational effects become
important.  This expectation is indeed realized in our numerical simulations,
justifying the assumption used in~\cite{Israel} to  analytically estimate the
effects of the scalar  field on the charged black hole interior.  We find a
thick (in terms of $r$)  layer where the presence of scalar matter produces
only slight deviations from  an exact \rn\ solution.   This effect is most
pronounced in the function $g$ at  late times;  near to the event horizon
$g\simeq -1$ as in \rn\  spacetime,  however it rapidly approaches zero beyond
this region.

\begin{figure}
\leavevmode
\vskip-6cm
\hbox{\epsfxsize=8cm \epsfysize=10cm {\epsffile{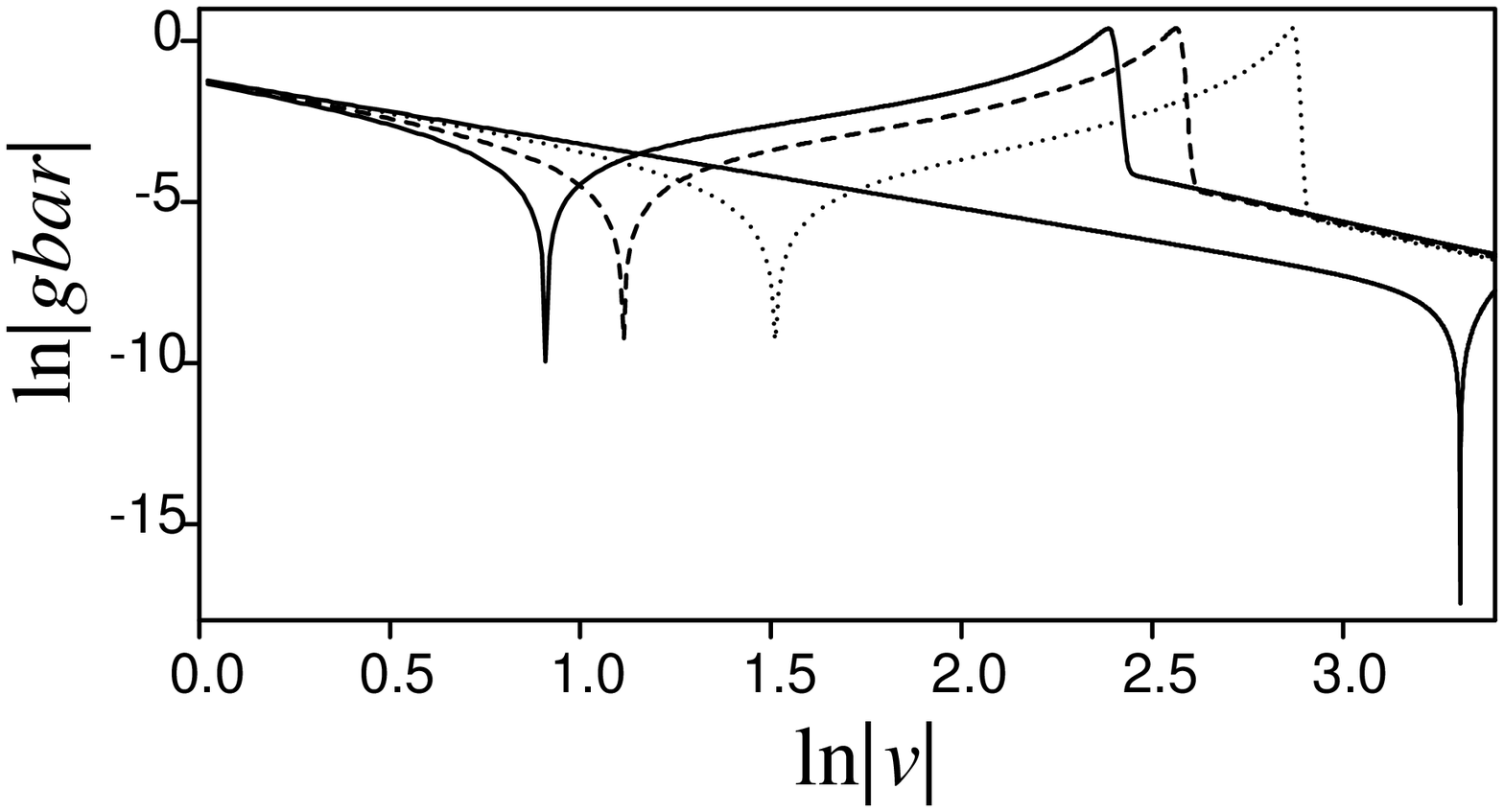}}}
\vskip-6cm
\hbox{\epsfxsize=8cm \epsfysize=10cm {\epsffile{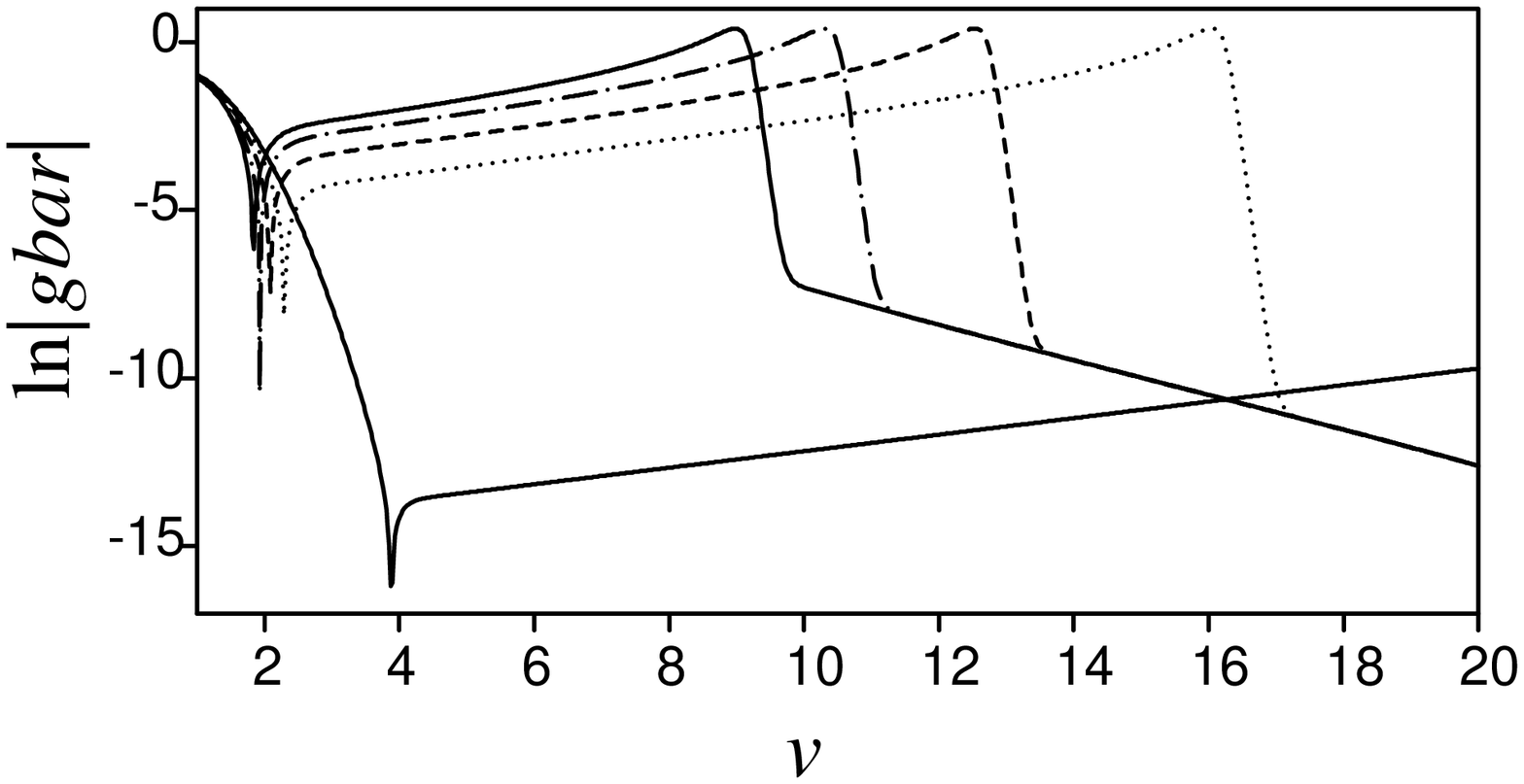}}}
\caption{\label{figure3} These plots show $\ln|\gbar|$ along a selection of
outgoing null rays which intersect the CH.  The top graph shows $\ln|\gbar|$
against $\ln|v|$ for the inverse power data in Table~I. The late time fall-off
is clearly a power law too.  The graph is $\ln|\gbar|$ against $v$ for the
exponential data.  The linear relation in this graph indicates an exponential
fall-off at late times.  The cusp in these figures corresponds to a change in
the sign of $\gbar$ where the outgoing null ray intersects the outer apparent
horizon of the black hole. }
\end{figure}

Figure~4 shows $\ln |g|$ plotted against advanced time along selected null
rays.    At late times $g\sim e^{-\gamma v}$,  where $\gamma$ depends only on
the charge $e$.  We find $\gamma \simeq \kappa_-$ as predicted by analytic
models~\cite{Ori:91}, where $\kappa_- = \sqrt{1 - e^2}/(1 - \sqrt{1 - e^2})^2$
is the surface gravity of the inner horizon of a static black hole with
equivalent mass and charge.  This result holds to within about 10\%.

While a suitable coordinate transformation can render the metric
non-singular~\cite{Ori:91}, the exponential decrease in $g$ is reflected in the
growth of curvature as the CH is approached.  In fact
$C^{\alpha\beta\gamma\delta}C_{\alpha\beta\gamma\delta} \simeq 48  m^2(u,v)/
r^6(u,v) \sim \gbar^2(u,v) e^{2\gamma v}$ along outgoing null rays. ($u$ labels
the outgoing null hypersurfaces on which we have examined the functions.)  For
a black hole of about one solar mass we have been able to follow the evolutions
to curvatures around Planck levels.

\begin{figure}
\leavevmode
\vskip-3.5cm
\hskip-0.5cm\hbox{\epsfxsize=8cm \epsfysize=10cm {\epsffile{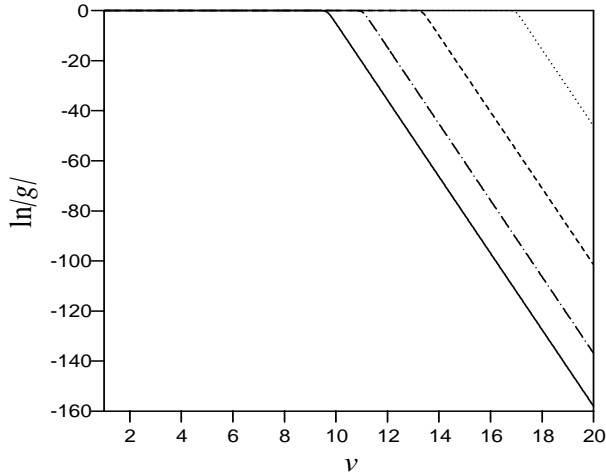}}}
\caption{\label{figure4} $\ln |g|$ against advanced time along the outgoing
null rays.   The asymptotic form $g \sim e^{-\gamma v}$ is in remarkable
agreement with predictions based on simplified models. }
\end{figure}

Analytic models have relied heavily on the results of linear perturbation
theory to provide information about energy fluxes along the CH.  Therefore,  we
have examined the flux of scalar field parallel to the CH in our solutions,
finding qualitative agreement with perturbative analyses.  It is
straightforward to see this remarkable feature directly from Raychaudhuri's
equation~(\ref{eq:9}) and the results in Table~\ref{table:1}.  If $l^{\alpha} =
dx^{\alpha}/dv $ is a lightlike generator of the outgoing characteristics then
the flux of scalar matter across these surfaces is
	\be
	{\cal F} := 4\pi T_{\alpha\beta}l^{\alpha} l^{\beta} =
			(d\phi/dv)^2  \; .
	\ee
Since $r \rightarrow {\rm constant}$ and $|g|$ decays exponentially at late
times, Eq.~(\ref{eq:9}) implies  that
	\be
	{\cal F} \simeq  A \gbar + B\, d\gbar/dv \ \  \mbox{\rm as}
	\ \  v\rightarrow\infty \; ,
	\ee
where $A$ and $B$ generally depend on the outgoing null surface considered.
Now,  if the flux across the event horizon at late times is ${\cal F}|_{\Gamma}
\sim v^{-p-1}$ then ${\cal F} \sim {\cal F}|_{\Gamma} \, v^{-\sigma}$ with
$\sigma >0$,  along null rays which intersect the CH.  The value of $\sigma$
(lying in the range $0.3< \sigma < 0.75$ for the cases we examined) depends on
the charge,  and differs from predicted values~\cite{Chandra,Israel} by more
than the numerical uncertainty.  Generally, if ${\cal F}|_{\Gamma}$  decays
faster than $\exp[-2\kappa_+v]$ for large $v$,  then ${\cal F} \sim  \exp[-2\mu
v]$  where $ \mu \simeq \kappa_+ = \sqrt{1 - e^2}/(1 + \sqrt{1 - e^2})^2$ (to
within less than 10\%).  The striking agreement with predictions of linearized
theory~\cite{Chandra} suggests that perturbative arguments~\cite{Ori:92,Israel}
are actually valid deep inside the black-hole core.

The most significant shortfall in our analysis is the location of the CH at
$v=\infty$.   So, how close to the CH do we really get?   Without the entire
solution (up to and including the CH) it is difficult to quantify this.
However,  in terms of a Kruskal-like coordinate $V = - e^{-\kappa_- v}$  which
goes to zero on the CH  we have reached values as small as $V = -10^{-150}$
during our integrations.   It might seem tempting to choose such an advanced
time coordinate  {\em ab initio},  so that the CH is located at a finite
coordinate distance ${\cal V}_{CH}$.  However a high price is paid for such a
choice;   in the new coordinates the initial data $\gbar|_{\Gamma}$ and
$g|_{\Gamma}$ become badly behaved as ${\cal V} \rightarrow {\cal V}_{CH}$
leading to significantly decreased integration times.

In conclusion,  we believe that the effects we have described above are
representative of the asymptotic structure of the true black hole interior and
that a null,  mass inflation singularity is present along the CH.
Furthermore, the null CH singularity is a precursor of the final spacelike
singularity deep within the black-hole core.  A detailed account of this work
is in preparation.

We are grateful to Ian Moss and Werner Israel for useful discussions.  This
research was supported in part by the EPSRC.

\narrowtext


\end{document}